\documentclass{article}

\usepackage[a4paper, total={5in, 8in}]{geometry}

\usepackage{graphicx}
\graphicspath{ {./images/} }

\renewcommand{\it}{\textit}

\title{\sc A Network-level View \\ of Author Influence}
\author{Henry Blanchette}
\date{May 1, 2019}

\begin{document}


\maketitle


\section{Introduction}
\label{sec:introduction}

The data set I used for this study is the record of papers published in a selection of computer systems conferences and a few control, non-computer systems conferences from 2017.
The original format was the raw PDF's of all of these papers.
During the summer of 2018, Eitan Frachtenberg
 and his research assistants extracted features from the PDF's including author names, paper titles, and bibliographies.
The second layer of data embellishment was matching authors with Google scholar accounts and personal survey results in order to attach author features such as email, institution, country, and gender.
Finally, a third layer of data embellishment was matching authors and papers to Semantic Scholar entries.
Nearly all the papers and authors (2,422 $\approx$ 99.1\%) from the original PDF's were found in Semantic Scholar.
Importantly, these Semantic Scholar identities significantly contributed to author disambiguation and retrieved the papers from previous years that were cited by the 2017 papers.

The stated purpose for my work on and analysis of this data was to look for patterns in the authors' and papers' individual-level features (from the embellished data) and network-level features.
I analyzed several different networks, focussing on the collaboration and citation connections between authors, papers, and conferences.
I found that the author collaboration network was particularly interesting because it displayed some significant, meaningful and separate correlations between individual-level features and network-level features.
These correlations were novel in the context of some more tradition \it{author reputation metrics}, which suggest that they measure an aspect of influence that is not well measured by the reputation metrics.

\section{Background}
\label{sec:background}

The study of social networks in general has detailed how networks can give rise to emergent, important features.
These sorts of features include: diameter, connected components, bridges, distance, and centrality.
I mainly focus on centrality, but the others (especially connected components and bridges) will also play well into the later analyses as complements in explaining centrality distributions.

A node's centrality in a network can have several different flavors as will be described, but abstract, centrality is a how \it{influential} a node is on the rest of the network (where ``influential'' is in terms of whatever the edges in the network represent).
There are a few specific measures of node centrality that have proven most useful for many network contexts.
They are \it{degree}, \it{eigenvector}, \it{betweenness}, and \it{closeness} centrality.

Degree centrality is the simplest.
The \it{degree} of a node is the number of edges that include the node (note that there can also concepts of in-degree and out-degree that only count edges of a certain orientation with respect to the given node).
Then the degree centrality of a node is defined just to be the node's degree.
Often, since centralities are considered at a network-level context, degree centrality is measured differently from degree only in being normalized over the degrees of the nodes of the entire network.
Degree centrality is a measure of how prolifically connected a node is to the network, as just a raw count of how many connections it has.
This measure is the most common used in network analysis, but it is also the most local of the centrality measures because it does not consider the arrangement of the network outside of each nodes' direct neighbors.

Eigenvector centrality is more complicated.
The concept is that it weights the centrality of each node by the centralities of its neighbors, after starting from some base ranking of nodes by degree.
In this way, it is closely related to PageRank [Rogers I.].
Formally, the eigenvector centrality of the nodes in a network is given by the vector $x$ that satisfies the equation $$ A x = \lambda x $$ where $A$ is the adjacency matrix of the network and $\lambda$ is an eigenvector with non-negative entries.
This requirement yields that there is a unique solution up to scaling. Therefor, eigenvector centrality is only meaningfully considered when normalized over the eigenvector centralities of all the nodes in the network.
Eigenvector centrality is a network-level-focussed measure because it weights heavily not just the direct neighbors of a node but also the neighbors of its neighbors and so on with diminishing effect.

Betweenness centrality measures how efficiently a node connects the network.
A given node $a$'s betweenness centrality is defined to be, over all pairs of nodes $c, d$ in the $a$'s connected component, the fraction of shortest paths between $c, d$ that include $a$.
Finally, the fraction is weighted by the size in nodes of $a$'s connected component.
Betweenness centrality is strongly connected to the concepts of connected components and bridges in a network. A connected component is a subset of the nodes of a network that have paths between any pair of them, but no paths to nodes outside the component.
Bridges are nodes that connect two would-be (non-trivial) connected components.
These bridges often have high betweenness because any path, and thus including the shortest ones, between nodes on opposite sides of the bridge must go through it.

Closeness centrality measures how close to the ``center of mass'' of the network a node is.
Closeness of a node is defined to be the connected-component-weighted reciprocal of its \textit{distance}, where the distance of a node is the mean length of the shortest paths between it and each other node in its connected component.
In a geographical way, many networks arrange themselves with a sort of planar flatness, so closeness centrality can be seen as how close a node literally is to the geographical center of the network.
Closeness is related to how involved a node is with the rest of its connected component as opposed to being involved with only a specific fringe (even if it is avidly involved which would lead to a high degree centrality).

All of these metrics are useful in the context of measuring ``influence'' but. of course, influence can be specified more clearly in specific situations.
For example, scientific collaboration analysts of computer science fields have used degree centrality along with correlations of node traits to explain why authors collaborate and how likely authors' with certain traits (e.g. gender, institution type) are to collaborate in the future [Ghiasi G.].
Along with degree, connections between nodes can be classified by other individual- and network-level features of the connected nodes [Ghiasi G].

More generally, the collaboration networks of scientific researchers have demonstrated that some particular features are important for predicting how likely some given authors are to collaborate. These features include: co-authorship distance in the author collaboration network, geographical closeness, cultural closeness, and likeness of research institution type (e.g. university or industry) [Knoke D, Yang S].
Some interesting results pointed to the conclusion that, perhaps, inter-disciplinary collaboration may be comparably important to intra-disciplinary collaboration [Knoke, Yang S].

\section{Hypotheses}
\label{sec:hypotheses}

My first inquiry was into the topical relationships of the conferences of the papers into the data.
Which conferences were most closely related to ``computer systems'' research, and how did the other distribute?
The data also contained papers from conferences that were identified as outside of computer systems in order to but the topical categorization into scale.
My hypothesis was that the typical intuitions about which computer systems conferences were most computer-systems-y would be right.
Of all people the scientists should know what kind of work they are doing, right?

My second inquiry was into the significance of typical reputation measures for authors, such as hindex and i10index, at the network level of analysis. More specifically, I predicted that reputation would correlate with the network-level sense of centrality that would be captured by some of the centrality metrics detailed in section \ref{sec:background}. The relationship between reputation and network-level node features should reveal something in regards to what the reputation metrics are and aren't good measures of.

\section{Conference Citations}
\label{sec:conference-citation}

In the conference citation network, each node represents a 2017 conference and each directed edge represents a citation by a paper in the source node's conference of a paper in the target node's conference.
Multiple with the same source and target combine into a single edge by summing weights (where weight is the number of citations that an edge represents).
These citations can go back in time however, as the conferences included in the data repeat regularly over the last couple decades.
Many citations were of conferences that were not present in the 2017 data; these were excluded from analysis.


\section{Paper Collaborations}
\label{sec:paper-collaboration}

In the paper collaboration network, each node represents a paper from the 2017 data and each undirected edge represents an author that is shared by each of the connected node's papers.
Multiple edges between the same two nodes combine into a single edge by summing weights (where weight is the number of shared authors that an edge represents).

\begin{figure}[h!]
  \centering
  \includegraphics
    [width=1.0\textwidth]
    {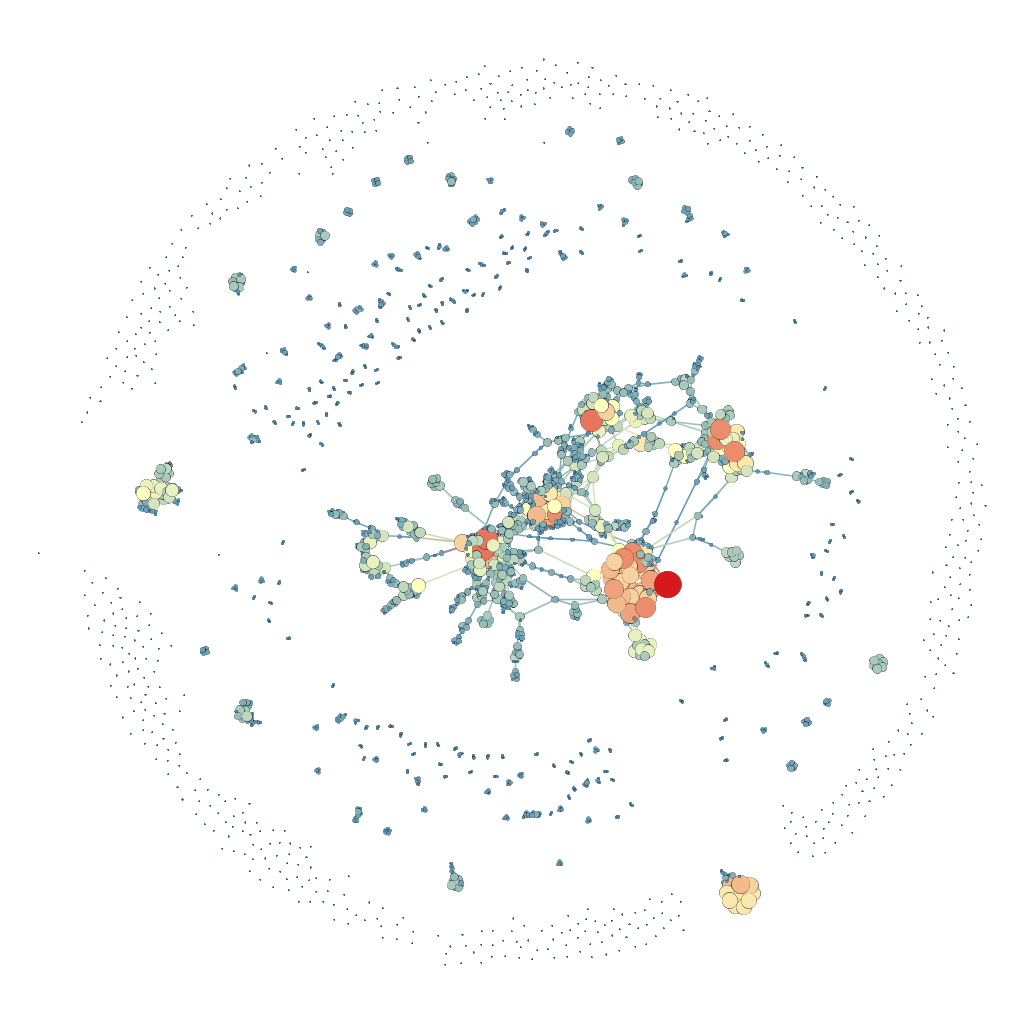}
  \caption{The paper collaboration network where the size and color of nodes correspond to their degree centrality. The color scale is from blue (low centrality) to white (middling centrality) to red (high centrality).}
  \label{fig:paper-collaboration-network-centrality-degree}
\end{figure}

As \ref{fig:paper-collaboration-network-centrality-degree} illustrates, this network is very clean visually.
It also gives a sense for the shape of the network.
There is a distinct, largest connected component that contains the vast majority of the highly degree-centric nodes.
Outside of this there are only a few other notably-large components, and only one of which has notably-high degree-centric nodes (located in the bottom-right of figure~\ref{fig:paper-collaboration-network-centrality-degree}).

I found, however, that for the purpose of analyzing the features of authors, this network was sub-optimal.
Grouping together authors by paper obscure what are likely important author attributes that are heterogeneous among the co-authors of a particular paper. Taking these considerations into account, I went on to create analyze the network in section~\ref{sec:author-collaboration}.

\section{Author Collaborations}
\label{sec:author-collaboration}

In the author collaboration network, each node represents an author and each undirected edge represents a paper from the 2017 data that the connected authors collaborated on.
Multiple edges between the same nodes are combined into a single edge by summing weights (where the weight of an edge is the number of co-authored papers between the connected nodes' authors).

\begin{figure}[h!]
  \centering
  \includegraphics
    [width=1.0\textwidth]
    {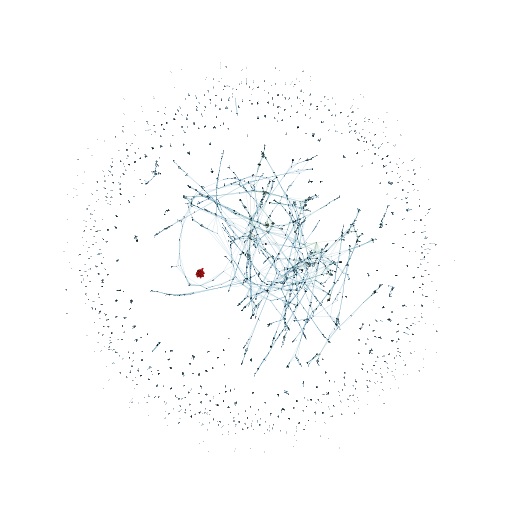}
  \caption{The author collaboration network where the size and color of nodes correspond to their degree centrality. The color scale is from blue (low centrality) to white (middling centrality) to red (high centrality).}
  \label{fig:authors-network-centrality-degree}
\end{figure}

As is immediately obvious from figure \ref{fig:authors-network-centrality-degree}, there is one particular group of nodes that skews the degree centrality of the whole network.
This fully-connected clique of authors are those that authored the paper.

\begin{figure}[h!]
  \centering
  \includegraphics
    [width=1.0\textwidth]
    {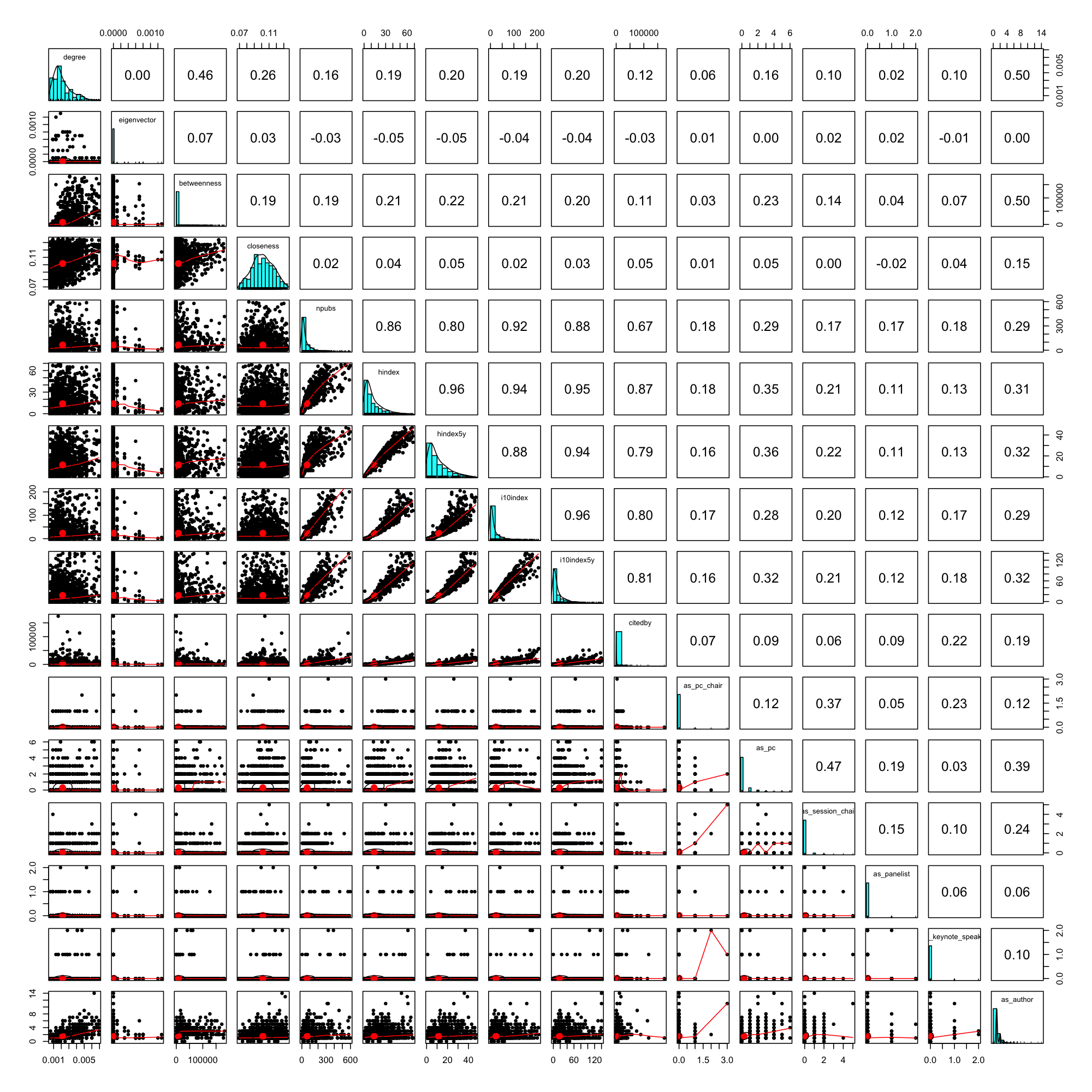}
  \label{fig:authors-feature-pairs-matrix}
  \caption{The author features' correlations matrix, including centralities.}
\end{figure}

\section{Reputation and Centrality}

Recall section \ref{sec:author-collaboration} where I listed and illustrated the network-level features of the author collaboration network.
In figure \ref{fig:authors-feature-pairs-matrix}, the network-level node centralities along with all other recorded author features are correlated against each other in pairs.
This correlations matrix yields some interesting results about how the different centrality distributions can be interpreted, but first to justify the pruning of some less-interesting correlations.

First of all, the following features did not correlate interestingly with any other features:
\begin{itemize}
\item as\_pc chair
\item as\_pc
\item as\_session\_chair
\item as\_panelist
\item as\_keynote\_speaker
\end{itemize}
This is likely because these specific measures turn out to be extremely bi-modally and coarsely distributed as opposed to the other author features.
If an author is a panelist in 2017 they only do it once or twice and the vast majority of authors are never panelists, and likewise for as\_pc\_chair, as\_pc, etc.

Secondly, the following features are extremely highly correlated (greater than 0.9) with each other and correlate highly similarly with every other feature: h10index, h10index5y, i10index, i10index5y.
Since each of these is a popular reputation metric for authors, this result is unsurprising; they are all attempting to measure roughly the same thing.
I chose h10index as a representative of this group so that they other may be omitted from further correlations matrices.

The most interesting results were that of how the different centrality metrics correlated and did not correlate with the author features that I expected to be related to centrality/influence in the collaborations network.
I analyzed each centrality metric individually.

The eigenvector centrality distribution was extremely bi-modal in a similar way to the as panelist feature, but without good reason to be so.
Eitan and I looked at some of the authors that obtained high centrality scores, but they did not end up passing the sniff test for relevance.
Since the distribution of these centralities was so incomparable to all other metrics and it seemed to elevate authors that did not have a clear reason for being considered so highly influential, I decided to omit this metric from further consideration.
Eigenvector centrality can be thought of as a crude version of PageRank, which clearly has very good use cases.
For example, it is accepted as well correlated with ``relatedness'' in the context of the network of web pages connected by hyperlinks.

However, the social network of author collaborations seems to be a case where this kind of centrality was not related to the concepts of influence or reputation.
Perhaps eigenvector centrality placed too much weight on certain connections because nodes in this network tend not to have more than a few edges.
Additionally, eigenvector centrality seemed to not be commonly used in the context of social networks in general so there may be a mismatch in setup of social networks and ``reference networks'' in regards to the ideal for eigenvector centrality.

The degree centrality distribution was one of the more immediately calculable and obvious metrics.
It is also one of the usual metrics considered in social network research.
Degree centrality correlated highest with betweenness ($0.46$) and as\_author ($0.50$).
It correlated negligibly ($< 0.3$) with all other author features.

The fact that degree and betweenness centralities correlate made sense because being connected to more nodes increases the chances that a node will end up on a shortest path.
However the correlation is not extremely high (it was still $< 0.6$), so there were clearly differences in what attributes many co-authors to an author and how exactly an author chooses their co-authors as to be most efficiently connected to the rest of the network.

The fact that degree and as\_author correlate made even more sense because authoring more papers usually involves more than just exactly the same people for each paper.
Again, the correlation is not extremeley high (it is still $< 0.6$), so there was a effect of authors ``in-collaborating'' with authors that they have already collaborated with during the year.
This was expected because authors are more likely to collaborate with other authors tehy already know, and one way to get to know a co-author is in fact by co-authoring with them.

The more unexpected results were the lack or correlation between degree centrality and each of npubs and hindex.
This is significant because it strongly suggested that the raw number of collaborations that an author does had little to do with either an author's reputation (hindex) or how likely they are to publish papers (npubs).
So collaborative-ness (in the degree centrality sense) is mostly orthogonal to reputation.

The betweenness centrality distribution has a large group hovering around 0 and an extremeley long tail of higher values.
The large group hovering around 0 was was all the authors that had only a couple of co-authors and were not part of a larger component.
These authors appear as the unconnected fringes of figure \ref{fig:authors-network-centrality-degree}.
However, even when only the distribution of betweenness in the largest component was considered there was a long tail.
Thus suggested that there is a fractal fringing of betweenness at varying levels of analysis (how many connected components are included in the analysis).
There were a few central hubs in the network that act as bridges between many other nodes.

Betweenness centrality correlated highest with degree centrality (0.46) and as\_author (0.5).
These results made sense for similar reasons as degree centrality.
Since correlation did not establish a ``causal'' relationship by itself, the correlation of betweenness and degree centralities can be explained in the opposite direction by the observation that an author's being more efficiently (in terms of their choice of co-authors) connected to the network increases the likelihood that that author has more co-authors.

Betweenness correlated negligibly ($< 0.23$) with all other features.
This is another surprising result alongside degree centrality's negligibly correlation with other features.
So this result likewise suggests that betweenness measures something orthogonal to the traditional measure of reputation.
Additionally, the non-correlation with npubs suggests that being efficiently connected to the network (betweenness) had little to do with how many papers an author previously wrote. In other words, the career output of an author does not predict how influential they are on the collaborations network.

Finally, the closeness centrality distribution was separated into two groups when considering the entire network. One group (the left-hand group of the betweenness distribution in figure \ref{fig:authors-feature-pairs-matrix}) was log-normal and the other group (the right-hand group of the betweenness distribution in figure \ref{fig:authors-feature-pairs-matrix} or the entire distribution of just the central component was normal.
Since component size was log-normally distributed, these distribution shapes suggested that closeness centrality was distributed normally in each component.
So furthermore the closeness seemed to be relatively randomly distributed; there were no clear center of influence in the network.
Rather, within each component (most evidently in the largest component) there were separate would-be subcomponents that are connected by highly between bridges.

Altogether these results suggested that centrality, in particular degree and betweenness centrality, measure a sense of author \textit{influence} in scholarly collaborations that were distinct from reputation or traditional measures of collaboration.
Not only did betweenness and degree centrality not correlate with other author features (other than as\_author), they did not correlate extremely with each other either.
The influence of an author on the collaborations of other authors thus seemed to be more nuanced than just local measures of collaborations counts or history of (successful) publications. What made an author an effective influence on the collaborations of their peers was a matter of traits mostly orthogonal to their own personal scholarly product.

\section*{References}

\noindent
Ghiasi G. \textit{On the Compliance of Women Engineers with a Gendered Scientific System}.

\noindent Rogers I. \textit{The Google PageRank Algorithm and How it Works}.\\https://www.cs.princeton.edu/~chazelle/courses/BIB/pagerank.htm.

\noindent
Knoke D. and Yang S. \textit{Social Network Analysis}.

\noindent
Sonnenwald D. \textit{Scientific Collaboration}.

\end{document}